\documentclass[a4paper]{article}
\usepackage[utf8]{inputenc}
\usepackage[english]{babel}
\usepackage{shortcuts}
\usepackage[matrix,arrow,curve]{xy}
\usepackage{hyperref}
\usepackage{wrapfig}
\usepackage{graphicx}
\usepackage{a4wide}
\usepackage{amsmath}
\usepackage{amsfonts}
\usepackage{theorem}

{\theorembodyfont{\rmfamily}
\newtheorem{definition}{Definition}
\newtheorem{example}{Example}
\newtheorem{remark}{Remark}
}

\newtheorem{proposition}[definition]{Proposition}

\newcommand{\strid}[1]{\scpdfinput{0.9}{#1.ps}}

\renewcommand{\leq}{\leqslant}
\newenvironment{rpar}{\begin{list}{}{\vspace{-2.7ex}\setlength{\leftmargin}{3ex}}\item[]}{\end{list}}

\newcommand{\D}{\mathcal{D}}

\newcommand{\Pol}{\category{Pol}}

\newcommand{\nPol}[1]{\textrm{$#1$-}\Pol}

\newcommand{\nCPol}[1]{\textrm{$#1$-}\category{CPol}}
\newcommand{\Net}[2]{\textrm{$#1$-}\category{Net}_{#2}}

\newcommand{\eqth}[1]{#1} 
\newcommand{\fcat}[2]{\mathcal{C}_{#1}(#2)} 
\newcommand{\fcomp}[1]{\overline{#1}} 

\newcommand{\Id}{\mathrm{Id}}

\newcommand{\Bij}{\category{Bij}}

\newcommand{\contexts}[1]{\mathcal{K}(#1)}

\newcommand{\fpoly}[1]{#1^*}

\newcommand{\becomes}{\mathop{:=}}

\title{Computing Critical Pairs \\in 2-Dimensional Rewriting Systems}
\author{Samuel Mimram\thanks{CEA, LIST, Point Courrier 94, 91191 Gif-sur-Yvette, France.}}

\hypersetup{
  pdftitle={\csname @title\endcsname},
  pdfauthor={Samuel Mimram},
  unicode=true,
  colorlinks=true,
  linkcolor=black,
  citecolor=black,
  urlcolor=black
}
\renewcommand{\C}{\mathcal{C}}

\begin{document}
\maketitle

\begin{abstract}
  \noindent Rewriting systems on words are very useful in the study of
  monoids. In good cases, they give finite presentations of the monoids,
  allowing their manipulation by a computer. Even better, when the presentation
  is confluent and terminating, they provide one with a notion of canonical
  representative for the elements of the presented monoid. Polygraphs are a
  higher-dimensional generalization of this notion of presentation, from the
  setting of monoids to the much more general setting of $n$-categories. Here,
  we are interested in proving confluence for polygraphs presenting
  $2$-categories, which can be seen as a generalization of term rewriting
  systems. For this purpose, we propose an adaptation of the usual algorithm for
  computing critical pairs. Interestingly, this framework is much richer than
  term rewriting systems and requires the elaboration of a new theoretical
  framework for representing critical pairs, based on contexts in compact
  $2$-categories.\footnote{This work was started while I was in the PPS team
    (CNRS -- Univ. Paris Diderot) and has been supported by the CHOCO (``Curry
    Howard pour la Concurrence'', ANR-07-BLAN-0324)
  French ANR project.}
\end{abstract}


Term rewriting systems have proven very useful to reason about terms modulo
equations. In some cases, the equations can be oriented and completed in a way
giving rise to a converging (\ie confluent and terminating) rewriting system,
thus providing a notion of canonical representative of equivalence classes of
terms. Usually, terms are freely generated by a \emph{signature}
$(\Sigma_n)_{n\in\N}$, which consists of a family of sets~$\Sigma_n$ of
generators of arity~$n$, and one considers \emph{equational theories} on such a
signature, which are formalized by sets of pairs of terms called
\emph{equations}. For example, the equational theory of monoids contains two
generators $m$ and $e$, whose arities are respectively~2 and~0, and three
equations
\[
m(m(x,y),z)=m(x,m(y,z))
\qquad
m(e,x)=x
\qtand
m(x,e)=x
\]
These equations, when oriented from left to right, form a rewriting system which
is converging. The termination of this system can be shown by giving an
interpretation of the terms in a well-founded poset, such that the rewriting
rules are strictly decreasing. Since the system is terminating,
the confluence can be deduced from the local confluence, which can itself be
shown by verifying that the five critical pairs
\[
m(m(m(x,y),z),t)
\qquad
m(m(e,x),y)
\qquad
m(m(x,e),y)
\qquad
m(m(x,y),e)
\qquad
m(e,e)
\]
are joinable and these critical pairs can be computed using a unification
algorithm. A more detailed presentation of term rewriting systems along with the
classic techniques to prove their convergence can be found
in~\cite{baader-nipkow:trat}.

As a particular case, when the generators of an equational theory are of arity
one, the category of terms modulo the congruence generated by the equations is a
monoid, with addition given by composition and neutral element being the
identity. A \emph{presentation} of a monoid~$(M,\times,1)$ is such an equational
theory, which is generating a monoid isomorphic to~$M$. For example the monoid
$\N/2\N$ is presented by the equational theory with only one generator~$a$, of
arity one, and the equation~$a(a(x))=x$. Presentations of monoids are
particularly useful since they can provide finite description of monoids which
may be infinite, thus allowing their manipulation with a computer. More
generally, with generators of any arity, equational theories give rise to
presentations of Lawvere theories~\cite{lawvere:phd}, which are cartesian
categories whose objects are the natural integers and such that product is given
on objects by addition: a signature namely generates such a category, whose
morphisms~\hbox{$f:m\to n$} are $n$-uples of terms with~$m$ free variables,
composition being given by substitution.

Term rewriting systems have been generalized by \emph{polygraphs}, in order to
provide a formal framework in which one can give presentations of any (strict)
$n$-category. We are interested here in adapting the classical technique to
study confluence of 3\nbd{}polygraphs, which give rise to presentations of
2\nbd{}categories, by computing their critical pairs. These polygraphs can be
seen as term rewriting systems improved on the following points:
\begin{itemize}
\item the variables of terms are simply typed (this can be thought as
  generalizing from a Lawvere theory of terms to any cartesian category of
  terms),
\item variables in terms cannot necessarily be duplicated, erased or swapped
  (the categories of terms are not necessarily cartesian but only monoidal),
\item and the terms can have multiple outputs as well as
multiple inputs.
\end{itemize}
Many examples of presentations of monoidal categories where studied by
Lafont~\cite{lafont:boolean-circuits}, Guiraud
\cite{guiraud:presentations-petri-nets, guiraud:three-dimensions-proofs} and the
author~\cite{mimram:phd, mimram:first-order-causality}. A fundamental example is
the 3-polygraph~$\eqth{S}$, presenting the monoidal category~$\Bij$ (the
category of finite ordinals and bijections). This polygraph has one generator
for objects~$1$, one generator for morphisms $\gamma:2\to 2$ (where $2$ is a
notation for $1\otimes 1$) and two equations
\begin{equation}
\label{eq:bij-eq}
  (\gamma\otimes 1)\circ(1\otimes\gamma)\circ(\gamma\otimes1)
  =
  (1\otimes\gamma)\circ(\gamma\otimes  1)\circ(1\otimes\gamma)
  \qqtand
  \gamma\circ\gamma=1\otimes 1
\end{equation}
where the morphism $1$ is a short notation for $\id_1$. That this polygraph is a
presentation of the category~$\Bij$ means that this category is isomorphic to
the free monoidal category containing an object~$1$ and a generator $\gamma$,
quotiented by the smallest congruence generated by the
equations~\eqref{eq:bij-eq}. This result can be seen as a generalization of the
presentation of the symmetric groups by transpositions. These equations can be
better understood with the graphical notation provided by \emph{string
  diagrams}, which is a diagrammatic notation for morphisms in monoidal
categories, introduced formally
in~\cite{joyal-street:geometry-tensor-calculus}. The morphism~$\gamma$ should be
thought as a device with two inputs and two outputs of type~$1$,
and the two equations~\eqref{eq:bij-eq} can thus be represented graphically by
\begin{equation}
  \label{eq:bij-string}
  \strid{gamma_yb_l}
  =
  \strid{gamma_yb_r}
  \qqtand
  \strid{gamma_sym_l}
  =
  \strid{gamma_sym_r}
\end{equation}
In this notation, wires represent identities (on the object $1$), horizontal
juxtaposition of diagrams corresponds to tensoring, and vertical linking of
diagrams corresponds to composition of morphisms. Moreover, these diagrams
should be considered modulo planar continuous deformations, so that the axioms
of monoidal categories are verified. These diagrams are conceptually important
because they allow us to see morphisms in monoidal categories either as
algebraic objects or as geometric objects (some sort of planar graphs). If we
orient both equations from left to right, we get a rewriting system which can be
shown to be convergent. It has the three following critical
pairs~\cite{lafont:boolean-circuits}:
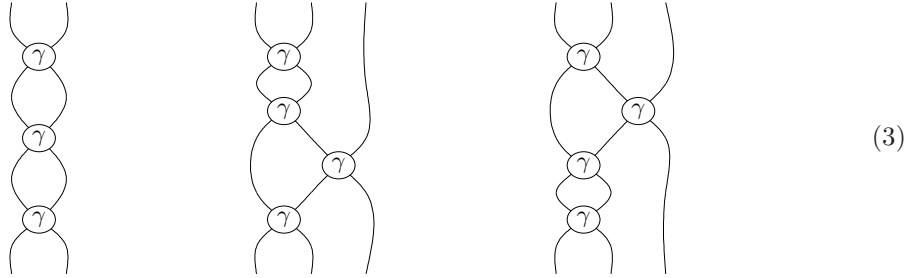
\begin{equation}
  \label{eq:ex-cp}
  \strid{gamma_cp_1}
  \qquad\qquad\qquad
  \strid{gamma_cp_2}
  \qquad\qquad\qquad
  \strid{gamma_cp_3}
\end{equation}
Moreover, for every morphism $\phi:1\otimes m\to 1\otimes n$, the morphism on
the left of~\eqref{eq:cp}
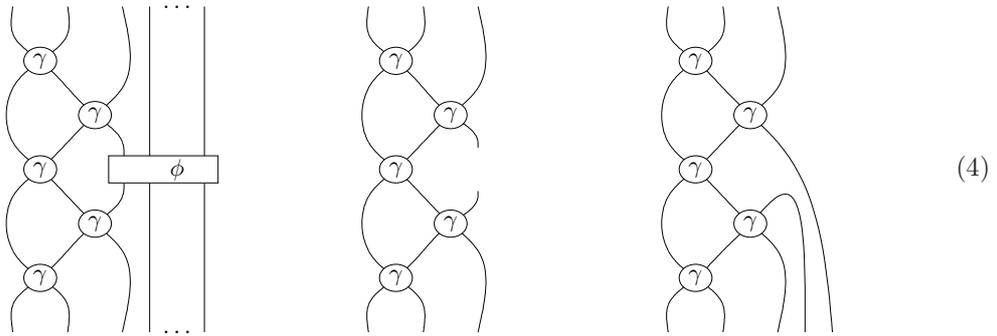
\begin{equation}
  \label{eq:cp}
  \strid{gamma_cp}
  \qquad\qquad
  \strid{gamma_cp_ctxt}
  \qquad\qquad\qquad
  \strid{gamma_cp_ctxt_compact}
\end{equation}
can be rewritten in two different ways, thus giving rise to an infinite number
of critical pairs for the rewriting system. This phenomenon was first observed
by Lafont~\cite{lafont:boolean-circuits} and later on studied by Guiraud and
Malbos~\cite{guiraud-malbos:higher-cat-fdt}. Interestingly, we can nevertheless
consider that there is a finite number of critical pairs if we allow ourselves
to consider the ``diagram'' on the center of~\eqref{eq:cp} as a critical
pair. Of course, this diagram does not make sense at first. However, we can give
a precise meaning to it if we embed our terms in a larger category, which is
compact: in such a category every object has a dual, which corresponds
graphically to having the ability to bend wires (see the figure on the
right). This observation was the starting point of this paper which is devoted
to formalizing these intuitions in order to propose an algorithm for computing
critical pairs in polygraphs.

We believe that this is a major area of higher-dimensional algebra where
computer scientists should step in: typical presentations of categories can give
rise to a very large number of critical pairs and having automated tools to
compute them seems to be necessary in order to push further the study of those
systems. The present paper constitutes a first step in this direction, by
defining the structures necessary to manipulate algorithmically the morphisms in
categories generated by polygraphs and by proposing an algorithm to compute the
critical pairs in polygraphic rewriting systems. Conversely, algebra provides
strong indications about technical choices that should be made in order to
generalize rewriting theory in higher dimensions. We have done our possible to
provide an overview of the theoretical tools used here, as well as intuitions
about them. A preliminary detailed version of this work is available
in~\cite{mimram:ccpp}.

We begin by recalling the definition of polygraphs, describe the categories they
generate, and formulate the unification problem in this framework using the
notion of context in a 2-category. Then, we show that 2-categories can be fully
and faithfully embedded into the free compact 2-category they generate, which
allows us to describe a unification algorithm for polygraphic rewriting systems.

\section{Presentations of 2-categories}
\label{sec:polygraphs}
Because of space limitations, we have to omit the basic definitions in category
theory and refer the reader to MacLane's reference book~\cite{maclane:cwm}. We
only recall that a \emph{2-category} is a generalization in dimension 2 of the
concept of category. It consists essentially of a class of \emph{$0$-cells} $A$,
a class of \emph{$1$\nbd{}cells} $f:A\to B$ (with 0-cells $A$ and $B$ as source
and target) and a class of \emph{$2$-cells}~$\alpha:f\To g:A\to B$ (with
parallel 1-cells $f:A\to B$ and~$g:A\to B$ as source and target), together with
a \emph{vertical composition}, which to every pair of 2-cells~$\alpha:f\To g$
and~$\beta:g\To h$ associates a 2-cell~$\beta\circ\alpha:f\To h$, and a
\emph{horizontal composition}, which to every pair of 2-cells~$\alpha:f\To g$
and~$\beta:h\To i$ associates a 2-cell~$\alpha\otimes\beta:(f\otimes
h)\To(g\otimes i)$, such that vertical and horizontal composition are
associative, admit neutral elements (the identities) and the \emph{exchange law}
is satisfied: for every four 2-cells
\[
\alpha:f\To f':A\to B,
\quad
\alpha':f'\To f'':A\to B,
\quad
\beta:g\To g':B\to C,
\quad
\beta':g'\To g'':B\to C
\]
the following equality holds
\begin{equation}
  \label{eq:exchange-law}
  (\alpha'\circ\alpha)\otimes(\beta'\circ\beta)
  \qeq
  (\alpha'\otimes\beta')\circ(\alpha\otimes\beta)
\end{equation}
as well as a nullary version of this law: $\id_{A\otimes B}=\id_A\otimes\id_B$
for every objects~$A$ and~$B$. In a $2$-category, two $n$-cells are
\emph{parallel} when they have the same source and the same target. We also
recall that two 0-cells~$A$ and~$B$ of a 2-category~$\C$, induce a
category~$\C(A,B)$, called \emph{hom-category}, whose objects are the 1-cells
$f:A\to B$ of~$\C$ and whose morphisms~$\alpha:f\To g$ are 2-cells of~$\C$,
composition being given by vertical composition. A~(strict) \emph{monoidal
  category} is a 2-category with exactly one 0-cell.

Polygraphs are algebraic structures which were introduced in their
2\nbd{}dimensional version by Street~\cite{street:limit-indexed-by-functors}
under the name \emph{computads}, generalized to higher dimensions by
Power~\cite{power:n-cat-pasting}, and independently rediscovered by
Burroni~\cite{burroni:higher-word}. We are specifically interested in
3-polygraphs, which give rise to presentations of 2-categories, and briefly
recall their definition here. This definition is a bit technical but
conceptually clear: it consists of sets of 0-, 1-, 2-generators for ``terms'',
each 2-generator having a list of 1-generators as source and as target, each
1-generator having itself a 0-generator as source and as target, together with a
set of equations which are pairs of terms (generated by the 2-generators).

Suppose that we are given a set~$E_0$ of \emph{0-generators}, such a set will be
called a \emph{0\nbd{}polygraph}. We write $\fpoly{E_0}=E_0$ and
$i_0:E_0\to\fpoly{E_0}$ the identity function.
A 1-polygraph on these generators is a graph, that is a diagram
$\vxym{\fpoly{E_0}&\ar@<-0.7ex>[l]_{s_0}\ar@<0.7ex>[l]^{t_0}E_1}$
in~$\Set$, with~$E_0^*$ as vertices, the elements of~$E_1$ being called
\emph{1-generators}. We can construct a free category on this graph: its
set~$\fpoly{E_1}$ of morphisms is the set of paths in the graph (identities are
the empty paths), the source $\fpoly{s_0}(f)$ (\resp target $\fpoly{t_0}(f)$) of
a morphism $f\in\fpoly{E_1}$ being the source (\resp target) of the path. If we
write~$i_1:E_1\to\fpoly{E_1}$ for the injection of the 1-generators into
morphisms of this category, which to every 1-generator associates the
corresponding path of length one, we thus get a diagram
\begin{equation}
  \label{eq:one-poly-free}
  \vxym{
    E_0\ar[d]_{i_0}&\ar@<-0.7ex>[dl]_{s_0}\ar@<0.7ex>[dl]^{t_0}E_1\ar[d]_{i_1}\\
    \fpoly{E_0}&\ar@<-0.7ex>[l]_{\fpoly{s_0}}\ar@<0.7ex>[l]^{\fpoly{t_0}}\fpoly{E_1}\\
  }
\end{equation}
in~$\Set$, which is commutative in the sense that~$\fpoly{s_0}\circ i_1=s_0$ and
$\fpoly{t_0}\circ i_1=t_0$. A \emph{2-polygraph} on this 1-polygraph consists of
a diagram
\begin{equation}
  \label{eq:two-poly}
  \vxym{
    E_0\ar[d]_{i_0}&\ar@<-0.7ex>[dl]_{s_0}\ar@<0.7ex>[dl]^{t_0}E_1\ar[d]_{i_1}&\ar@<-0.7ex>[dl]_{s_1}\ar@<0.7ex>[dl]^{t_1}E_2\\
    \fpoly{E_0}&\ar@<-0.7ex>[l]_{\fpoly{s_0}}\ar@<0.7ex>[l]^{\fpoly{t_0}}\fpoly{E_1}\\
  }
\end{equation}
in~$\Set$, such that $\fpoly{s_0}\circ s_{1}=\fpoly{s_0}\circ t_{1}$ and
$\fpoly{t_0}\circ s_{1}=\fpoly{t_0}\circ t_{1}$. The elements of~$E_2$ are
called \emph{2-generators}. Again we can generate a free 2-category on this
data, whose underlying category is the category generated
in~\eqref{eq:one-poly-free} and which has the 2-generators as morphisms. If we
write~$\fpoly{E_2}$ for its set of morphisms and \hbox{$i_2:E_2\to\fpoly{E_2}$}
for the injection of the 2-generators into morphisms, we thus get a diagram
\begin{equation}
  \label{eq:two-poly-free}
  \vxym{
    E_0\ar[d]_{i_0}&\ar@<-0.7ex>[dl]_{s_0}\ar@<0.7ex>[dl]^{t_0}E_1\ar[d]_{i_1}&\ar@<-0.7ex>[dl]_{s_1}\ar@<0.7ex>[dl]^{t_1}E_2\ar[d]_{i_2}\\
    \fpoly{E_0}&\ar@<-0.7ex>[l]_{\fpoly{s_0}}\ar@<0.7ex>[l]^{\fpoly{t_0}}\fpoly{E_1}&\ar@<-0.7ex>[l]_{\fpoly{s_1}}\ar@<0.7ex>[l]^{\fpoly{t_1}}\fpoly{E_2}\\
  }
\end{equation}
We can now formulate the definition of 3-polygraphs as follows.

\begin{definition}
  \label{def:three-poly}
  A \emph{3-polygraph} consists of a diagram
  \begin{equation}
    \label{eq:three-poly}
    \vxym{
      E_0\ar[d]_{i_0}&\ar@<-0.7ex>[dl]_{s_0}\ar@<0.7ex>[dl]^{t_0}E_1\ar[d]_{i_1}&\ar@<-0.7ex>[dl]_{s_1}\ar@<0.7ex>[dl]^{t_1}E_2\ar[d]_{i_2}&\ar@<-0.7ex>[dl]_{s_2}\ar@<0.7ex>[dl]^{t_2}E_3\\
      \fpoly{E_0}&\ar@<-0.7ex>[l]_{\fpoly{s_0}}\ar@<0.7ex>[l]^{\fpoly{t_0}}\fpoly{E_1}&\ar@<-0.7ex>[l]_{\fpoly{s_1}}\ar@<0.7ex>[l]^{\fpoly{t_1}}\fpoly{E_2}\\
    }
  \end{equation}
  (where $\fpoly{E_i}$, $\fpoly{s_i}$ and $\fpoly{t_i}$ are freely
  generated as previously explained), such that
  \[
  \fpoly{s_i}\circ s_{i+1}=\fpoly{s_i}\circ t_{i+1}
  \qtand
  \fpoly{t_i}\circ s_{i+1}=\fpoly{t_i}\circ t_{i+1}
  \]
  for $i=0$ and $i=1$, together with a structure of 2-category on the 2-graph
  \[
  \vxym{
    \fpoly{E_0}&\ar@<-0.7ex>[l]_{\fpoly{s_0}}\ar@<0.7ex>[l]^{\fpoly{t_0}}\fpoly{E_1}&\ar@<-0.7ex>[l]_{\fpoly{s_1}}\ar@<0.7ex>[l]^{\fpoly{t_1}}\fpoly{E_2}
  }
  \]
\end{definition}
Again, a 3-polygraph freely generates a 3-category~$\C$ whose underlying
2\nbd{}category is the underlying 2-category of the polygraph and whose 3-cells
are generated by the 3-generators of the polygraph. A quotient
2-category~$\tilde\C$ can be constructed from this 2-category: it is defined as
the underlying 2-category of~$\C$ quotiented by the congruence identifying two
2-cells whenever there exists a 3\nbd{}cell between them in~$\C$. A
3-polygraph~$P$ \emph{presents} a 2\nbd{}category~$\D$ when~$\D$ is isomorphic
to the 2\nbd{}category~$\tilde\C$ induced by the polygraph~$P$. In this sense,
the underlying 2-polygraph of a 3-polygraph is a \emph{signature} generating
terms which are to be considered modulo the \emph{equations} described by the
3-generators; these equations~$r\in E_3$ being oriented, they will be called
\emph{rewriting rules}, the source~$s_2(r)$ (\resp the target~$t_2(r)$) being
the \emph{left member} (\resp \emph{right member}) of the rule. A polygraph is
\emph{finite} when all the sets~$E_i$ are; in the following, we only consider
such polygraphs.

A \emph{morphism of polygraphs} $F=(F_0,F_1,F_2,F_3)$ between two
3-polygraphs~$P$ and~$Q$ consists of functions $F_i:E_i^{P}\to E_i^{Q}$, such
that the obvious diagrams commute (for example, for every~$i$, $s_i^{Q}\circ
F_{i+1}=\fpoly{F_i}\circ s_i^{P}$, where
$\fpoly{F_i}:\fpoly{{E_i^{P}}}\to\fpoly{{E_i^{Q}}}$ is the monoid morphism
induced by~$F_i$). We write $\nPol{n}$ for the category of $n$-polygraphs (this
construction can be carried on to any dimension~$n\in\N$ but we will only
consider cases with~$n\leq 3$). These categories have many nice properties,
amongst which being cocomplete. The free $n$\nbd{}category generated by an
$n$-polygraph~$P$ is denoted~$\fcat{n}{P}$. Given an integer~$k\leq n$, we write
\hbox{$U_k:\nPol{n}\to\nPol{k}$} for the forgetful functor which simply forgets
about the sets of generators of dimension higher than~$k$. This functor admits a
left adjoint \hbox{$F_n:\nPol{k}\to\nPol{n}$} which adds empty sets of
generators of dimension higher than~$k$.
We sometimes leave implicit the inclusion of~$\nPol{k}$ into $\nPol{n}$ induced
by~$F_n$.

\begin{example}
  \label{ex:theory-sym}
  The theory of \emph{symmetries} mentioned in the introduction is the
  polygraph~$S$ whose generators are
  \[
  \begin{array}{l}
    E_0=\{*\}
    \qquad\qquad
    E_1=\{1:*\to *\}
    \qquad\qquad
    E_2=\{\gamma:1\otimes 1\To 1\otimes 1\}
    \\
    E_3=\{y:(\gamma\otimes 1)\circ(1\otimes\gamma)\circ(\gamma\otimes 1)\TO(1\otimes\gamma)\circ(\gamma\otimes 1)\circ(1\otimes\gamma),\ s:\gamma\circ\gamma\TO 1\otimes 1\}
  \end{array}
  \]
\end{example}

\begin{example}
  \label{ex:theory-monoids}
  The theory of \emph{monoids} is the polygraph~$\eqth{M}$ defined by
  \[
  \begin{array}{l}
    E_0=\{*\}
    \qquad\qquad
    E_1=\{1:*\to *\}
    \qquad\qquad
    E_2=\{\mu:1\otimes 1\To 1,\ \eta:*\To 1\}
    \\
    E_3=\{a:\mu\circ(\mu\otimes 1)\TO \mu\circ(1\otimes\mu),\ l:\mu\circ(\eta\otimes 1)\TO 1,\ r:(1\otimes\eta)\to 1\}
  \end{array}
  \]
  This polygraph presents the augmented simplicial category (the category of
  finite ordinals and non-decreasing functions).
\end{example}



\section{Formal representation of free 2-categories}
\label{sec:free-cat}
The definition of 3-polygraphs involves the construction of free categories and
free 2\nbd{}categories, which are abstractly defined in category theory by
universal constructions. Here, we need a more concrete representation of these
mathematical objects. As already mentioned, the free
category~\eqref{eq:one-poly-free} on a graph is easy to describe: its objects
are the vertices of the graph and morphisms are paths of the graph with
composition given by concatenation. However, describing the free 2\nbd{}category
on a 2\nbd{}polygraph in an effective way (which can be implemented) is much
less straightforward. Of course, following the definition given in
Section~\ref{sec:polygraphs}, one could describe the 2-cells of this 2-category
as formal vertical and horizontal compositions of 2-generators up to a
congruence imposing associativity and absorption of units for both compositions
and the exchange law~\eqref{eq:exchange-law}. However, given an object~$A$ in a
2-category~$\C$ and two 2-cells~$\alpha,\beta:\id_A\To\id_A:A\to A$ of this
category, the equality~$\alpha\otimes\beta=\beta\otimes\alpha$ can be deduced
from the following sequence of equalities: {\small
\[
\alpha\otimes\beta
=
(\id_A\circ\alpha)\otimes(\beta\circ\id_A)
=
(\id_A\otimes\beta)\circ(\alpha\otimes\id_A)
=
(\beta\otimes\id_A)\circ(\id_A\otimes\alpha)
=
(\beta\circ\id_A)\otimes(\id_A\circ\alpha)
=
\beta\otimes\alpha
\]}%
It requires inserting and removing identities, and using the exchange law in
both directions. So, it seems to be very hard to find a generic way to handle
formal composites of generators modulo the congruence described above. We will
therefore define an alternative construction of these morphisms which doesn't
require such a quotienting.

Consider the morphism~$\gamma\circ\gamma:(1\otimes 1)\To(1\otimes 1):*\to *$ in
the theory~$S$ of symmetries (Example~\ref{ex:theory-sym}), depicted on the left
of~\eqref{eq:cnet}:
\begin{equation}
  \label{eq:cnet}
  \strid{cnet_ex_l}
  \qquad\qquad\qquad\qquad
  \strid{cnet_ex_r}
\end{equation}
Graphically, in this morphism, the two 2-cells are~$\gamma$, wires are typed by
the 1-cell~$1$ and regions of the plane are typed by the 0-cell~$*$. Now, if we
give a different name to each \emph{instance} of a generator used in this
morphism, for example by numbering them as in the right of~\eqref{eq:cnet}, the
morphism itself can be described as the 2-polygraph~$P$ defined by
\[
E_0=\{*_0,\ldots,*_4\}
\qquad\qquad
E_1=\{1_0:*_1\to *_0,\ 1_1:*_0\to *_2,\ldots,\ 1_5:*_4\to *_2\}
\]
and
\[
E_2=\{\gamma_0:1_0\otimes 1_1\To 1_2\otimes 1_3,\ \gamma_1:1_2\otimes 1_3\To 1_4\otimes 1_5\}
\]
together with a function~$\ell$ which to every $i$-generator of this polygraph
associates a label, which is an $i$-generator of~$S$, so that~$\ell:P\to S$ is a
morphism of polygraphs ($\ell$ is defined by \hbox{$\ell(*_i)=*$},
\hbox{$\ell(1_i)=1$} and $\ell(\gamma_i)=\gamma$). Formulated in categorical
terms, $(P,\ell)$ is an object in the slice
category~$\slicecat{\nPol{2}}{U_2(S)}$. Of course, the naming of the instances
of the generators occurring in nets is arbitrary, so we have to consider these
labeled polygraphs up to bijections, which correspond to injective renaming of
instances. Notice that not every such labeled polygraph is the representation of
a morphism: we need an inductive construction of those (it seems to be difficult
to give a direct characterization of the suitable polygraphs).

Based on these ideas, we describe the category generated by a polygraph~$S$ as a
category whose cells are polygraphs labeled by~$S$. We suppose fixed a signature
2\nbd{}polygraph~$S$ and write~$S_i$ for $U_i(S)$. This is a generalization of
the constructions of labeled transition systems, and is reminiscent of pasting
schemes~\cite{power:n-cat-pasting} and of proof-nets, which is why we call them
\emph{polygraphic nets} (or \emph{nets} for short).

The category of \emph{$0$-nets}~$\Net{0}{S_0}$ on the $0$-polygraph~$S_0$ is the
full subcategory of \hbox{$\slicecat{\nPol{0}}{S_0}$} whose objects are
$0$-polygraphs with exactly one $0$-cell, labeled by~$S_0$. Concretely, its
objects are pairs $(n,A)$, often written~$A_n$, where~$n$ is the \emph{name} of
the instance (an integer for example) and~$A$ an element of $E_0^{S_0}$, called
its \emph{label}, and there is a morphism between two objects whenever they have
the same label (all those morphisms are invertible).
The category of \emph{$1$-nets}~$\Net{1}{S_1}$ is the smallest category whose
objects are the $0$\nbd{}nets~$A_i$, whose morphisms $(s^f,f,t^f):A_i\to B_j$
are triples consisting of a $1$-polygraph~$f$ labeled by~$S_1$ (\ie an object
in~$\slicecat{\nPol{1}}{S_1}$) and two morphisms of labeled polygraphs
$s^f:A_i\to f$ and $t^f:B_j\to f$, called \emph{source} and \emph{target}, which
are either a $1$\nbd{}polygraph~$f$ such that $E_0^f=\{A_i,B_j\}$ and~$E_1^f$
contains only one $1$-cell~$n\in\N$ with~$A_i$ as source and~$B_j$ as target
(and the obvious injections for~$s^f$ and $t^f$), or $A_i=B_j$, $f=A_i$ and
$s^f=t^f=\id_{A_i}$ (this is the identity on~$A_i$), or a composite~$f\otimes
g:A_i\to B_j$ of two morphisms~$f:A_i\to C_k$ and~$g:C_k\to B_j$. Here, the
composite of two such morphisms is defined as the pushout of the
diagram~$\vxym{f&\ar[l]_{t^f}C_k\ar[r]^{s^g}&g}$, that is the disjoint union of
the polygraphs~$f$ and~$g$ quotiented by a relation identifying the~$0$-cell
in~$C_k$ in the two components of the union.

\begin{example}
  If~$S$ is the polygraph of symmetries, the composite of the two morphisms
  \hbox{$f:*_0\to *_1$} and \hbox{$g:*_1\to *_2$} defined by
  \[
  E_0^f=\{*_0,\ *_1\}
  \ \ 
  E_1^f=\{1_0:*_0\to *_1\}
  \ \ 
  E_0^g=\{*_0,\ *_1,\ *_2\}
  \ \ 
  E_1^g=\{1_1:*_1\to *_0,\ 1_0:*_0\to *_2\}
  \]
  is the morphism $h=f\otimes g$ such that
  \[
  E_0^h=\{*_0,\ldots,*_3\}
  \qtand
  E_1^h=\{1_0:*_0\to *_1,\ 1_1:*_1\to *_3,\ 1_2:*_3\to *_2\}
  \]
  Graphically,
  \[
  \vxym{
    \ast_0\ar[r]^{1_0}&\ast_1
  }
  \quad\otimes\quad
  \vxym{
    \ast_1\ar[r]^{1_1}&\ast_0\ar[r]^{1_0}&\ast_2
  }
  \qeq
  \vxym{
    \ast_0\ar[r]^{1_0}&\ast_1\ar[r]^{1_1}&\ast_3\ar[r]^{1_2}&\ast_2
  }
  \]
\end{example}

\noindent Since composition is defined by a pushout construction, it involves a
renaming of some instances (it is the case in the example above) and this
renaming is arbitrary. So, composition is not strictly associative but only
associative up to isomorphism of polygraphs. Therefore, what we have built is
not precisely a category but only a bicategory: this is a well-known fact, this
construction being a particular instance of the general construction of cospan
bicategories. We can iterate this construction one step further and define the
tricategory (that is a 2-category whose compositions are associative up to
isomorphism) of \emph{$2$-nets}~$\Net{2}{S}$ as the smallest tricategory whose
$0$-cells are $0$-nets~$A_i$, whose $1$-cells~$f:A_i\to B_j$ contain
$1$\nbd{}nets, and whose $2$-cells $\alpha:f\To g$ are
triples~$(s^\alpha,\alpha,t^\alpha)$, consisting of a $2$-polygraph~$\alpha$
labeled by~$S$ and two morphisms of labeled polygraphs~$s^\alpha:f\to\alpha$
and~$t^\alpha:g\to\alpha$, containing all the $2$-polygraphs with one
$2$-generator $n\in\N$ whose source~$f=s_1^\alpha(n)$ and
target~$g=t_1^\alpha(n)$ are $1$-nets which are ``disjoint'' in the sense they
only have their own source and target as common generators, with the obvious
injections for~$s^\alpha$ and~$t^\alpha$. Moreover, we requires this tricategory
to contain identities and to be closed under both vertical and horizontal
compositions, which are defined by pushout constructions in a way similar to
$1$-nets. If we quotient this tricategory and identify cells which are
isomorphic labeled polygraphs, we get a proper $2$\nbd{}category, that we still
write~$\Net{2}{S}$.

\begin{proposition}
  The $2$-category~$\Net{2}{S}$ described above is equivalent to the free
  category generated by the $2$-polygraph~$S$.
\end{proposition}

This construction has the advantage to be simple to implement and manipulate: we
have for example given the data needed to describe the morphism~\eqref{eq:cnet}.

\section{Critical pairs in polygraphs}
\label{sec:cp}
In order to formalize the notion of critical pair for a polygraph, we need to
formalize first the notion of context of a morphism in the
$2$-category~$\fcat{2}{S}$ generated by a $2$-polygraph~$S$, which may be
thought as a $2$-cell with multiple typed ``holes''. These contexts have
multiples ``inputs'' (one for each hole) and will therefore organize into a
multicategory, which is a notion generalizing categories in the sense that
morphisms~$f:(A_1,\ldots,A_n)\to A$ have one output of type~$A$, and \emph{a
  list} of inputs of type~$A_i$ instead of only one input. Composition is also
generalized in the sense that we compose such a morphism~$f$ with~$n$
morphisms~$f_i$ with~$A_i$ as target, what we
write~$f\circ(f_1,\ldots,f_n)$. Multicategories should moreover have
identities~$\Id_A:(A)\to A$ and satisfy coherence
axioms~\cite{leinster:higher-operads}.

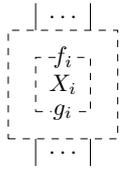
\begin{wrapfigure}{l}{16mm}
  \vspace{-4ex}
  \begin{center}
    $\strid{hole}$
  \end{center}
  \vspace{-4ex}
\end{wrapfigure}
Suppose that we are given a signature $2$-polygraph~$S$. Suppose moreover that
we are given a list of~$n$ pairs of parallel $1$\nbd{}cells $(f_i,g_i)$ in the
category generated by the $1$\nbd{}polygraph~$U_1(S)$. We
write~\hbox{$S[X_1:f_1\To g_1,\ldots,X_n:f_n\To g_n]$}, for the polygraph
obtained from~$S$ by adding $X_1,\ldots,X_n$ as $2$-generators, with~$f_i$ as
the source and~$g_i$ as the target of~$X_i$ (we suppose that the~$X_i$ were not
already present in the $2$-generators of~$S$). The~$X_i$ should be thought as
typed variables for $2$-cells and we can easily define a notion of
\emph{substitution} of a variable~$X_i:f_i\To g_i$ by a 2\nbd{}cell
$\alpha:f_i\To g_i$ in a 2\nbd{}cell of the $2$-category generated
by~$S[X_1:f_1\To g_1,\ldots,X_n:f_n\To g_n]$.

Given a signature~$S$, we build a multicategory~$\contexts{S}$ whose objects are
pairs~$(f,g)$ of parallel 1\nbd{}cells in the 2-category generated by~$S$ and
whose morphisms \hbox{$K:((f_1,g_1),\ldots,(f_n,g_n))\to(f,g)$}, called
\emph{contexts}, are the 2-cells~$\alpha:f\To g$ in the 2-category which is
generated by the polygraph \hbox{$S[X_1:f_1\To g_1,\ldots,X_n:f_n\To g_n]$},
which are \emph{linear} in the sense that each of the variables~$X_i$ appears
exactly once in the morphism~$\alpha$. Composition in this multicategory is
induced by the substitution operation. This multicategory can be canonically
equipped with a structure of symmetric multicategory, which essentially means
that, for every permutation~$\sigma$ on $n$ elements, the sets of morphisms of
type \hbox{$((f_1,g_1),\ldots,(f_n,g_n))\to(f,g)$} is isomorphic to the set of
morphisms of type
\hbox{$((f_{\sigma(1)},g_{\sigma(1)}),\ldots,(f_{\sigma(n)},g_{\sigma(n)}))\to(f,g)$}
in a coherent way. Any $2$-cell~$\alpha:f\To g$ in the $2$-category generated
by~$S$, can be seen as a nullary context of type~$()\to(f,g)$ that we still
write~$\alpha$.
A concrete and implementable definition of the multicategory~$\contexts{S}$ of
contexts of~$S$ can be given by adapting the construction of polygraphic nets
given in the previous section.

This construction enables us to reformulate usual notions of rewriting theory in
our framework as follows. We suppose fixed a rewriting system given by a
3-polygraph~$R$. We write~$S=U_2(R)$ for the underlying signature of~$R$
and~$\C$ for the 2-category it generates.
\begin{definition}
  \label{def:unifier}
  A \emph{unifier} of two 2-cells
  \[
  \alpha_1:f_1\To g_1
  \qtand
  \alpha_2:f_2\To g_2
  \]
  in $\C$ is a pair of cofinal unary contexts
  \[
  K_1:((f_1,g_1))\to(f,g)
  \qtand
  K_2:((f_2,g_2))\to(f,g)
  \]
  such that~$K_1\circ(\alpha_1)=K_2\circ(\alpha_2)$. A unifier is a \emph{most
    general unifier} when it is
  \begin{itemize}
  \item \emph{non-trivial}: there exists no binary context
    $K:((f_1,g_1),(f_2,g_2))\to(f,g)$ which satisfies
    \hbox{$K_1=K\circ(\Id_{(f_1,g_1)},\alpha_2)$}
    and~$K_2=K\circ(\alpha_1,\Id_{(f_2,g_2)})$. Informally, the
    morphisms~$\alpha_1$ and~$\alpha_2$ should not appear in disjoint positions
    in the morphism $K_1\circ(\alpha_1)=K_2\circ(\alpha_2)$.
  \item \emph{minimal}: for every unifier $K_1',K_2'$ of~$\alpha_1$
    and~$\alpha_2$, such that $K_1=K_1''\circ K_1'$ and $K_2=K_2''\circ K_2'$,
    for some contexts~$K_1''$ and~$K_2''$, the contexts~$K_1''$ and~$K_2''$
    should be invertible.
  \end{itemize}
\end{definition}

\noindent
\begin{remark}
  \label{rem:net-unifier}
  If we write~$\alpha=K_1\circ(\alpha_1)=K_2\circ(\alpha_2)$ and represent the
  2-cells~$\alpha_1$, $\alpha_2$ and~$\alpha$ by 2-nets, the fact that~$\alpha$
  is a unifier of the morphisms means that there exist two injective morphisms
  of labeled polygraphs~$i_1:\alpha_1\to\alpha$ and~$i_2:\alpha_2\to\alpha$, and
  the non-triviality condition means that there exists at least one 2-generator
  which is both in the image of~$i_1$ and~$i_2$.
\end{remark}

\noindent
For example, the last two morphisms of~\eqref{eq:ex-cp} are both unifiers of the
left members of the rules~\eqref{eq:bij-string}. By extension, a unifier of two
$3$-generators $r_1:\alpha_1\TO\beta_1$ and $r_2:\alpha_2\TO\beta_2$ of~$R$ is a
unifier of their sources~$\alpha_1$ and~$\alpha_2$. A \emph{critical pair}
$(K_1,r_1,K_2,r_2)$ consists of a pair of 3-generators $r_1,r_2$ and a most
general unifier~$K_1,K_2$ of those.

\begin{remark}
  \label{rem:unif-contexts}
  In Definition~\ref{def:unifier}, the 2-cell $\alpha_1$, can be seen as a
  context~$\alpha_1:()\to(f_1,g_1)$ in~$\contexts\C$, and similarly
  for~$\alpha_2$. In fact, the notion of \emph{unifier} can be generalized to
  any pair of morphisms in the multicategory~$\contexts\C$.
\end{remark}

A 2-cell $\alpha:f\To g$ \emph{rewrites} to a 2-cell~$\beta:f\To g$, by a
3-generator \hbox{$r:\alpha'\TO\beta':f'\To g'$}, when there exists a
context~$K:((f',g'))\to(f,g)$ such that~$\alpha=K\circ\alpha'$ and
\hbox{$\beta=K\circ\beta'$}. In this case, we write~$\alpha\TO^{K,r}\beta$. The
rewriting system~$R$ is \emph{terminating} when there is no infinite
sequence~$\alpha_1\TO^{K_1,r_1}\alpha_2\TO^{K_2,r_2}\ldots$.
A \emph{peak} is a triple $(\alpha_1,r_1,\alpha,r_2,\alpha_2)$, where $\alpha$,
$\alpha_1$ and $\alpha_2$ are $2$\nbd{}cells and~$r_1$ and~$r_2$ are
3\nbd{}generators, such that~$\alpha\TO^{K_1,r_1}$ and
\hbox{$\alpha\TO^{K_2,r_2}\alpha_2$}. In particular, with the notations of
Definition~\ref{def:unifier}, every critical pair induces a peak
\hbox{$(K_1\circ(\beta_1),r_1,K_1\circ(\alpha_1),r_2,K_2\circ(\beta_2))$}. A
peak is \emph{joinable} when there exist a $2$-cell~$\beta$ and $3$-cells
$\rho_1:\alpha_2\TO\beta$ and~$\rho_2:\alpha_2\TO\beta$. A rewriting system is
\emph{locally} confluent if every peak is joinable. Newman's Lemma is valid for
$3$-polygraphs~\cite{guiraud-malbos:higher-cat-fdt}:
\begin{proposition}
  A terminating rewriting system is confluent if it is locally confluent.
\end{proposition}
\noindent Moreover, local confluence can be tested using critical pairs:
\begin{proposition}
  A rewriting system is locally confluent if all its critical pairs are
  joinable.
\end{proposition}
\noindent So, in order to test whether a terminating polygraphic rewriting
system is confluent, it would be tempting to compute all its critical pairs and
test whether they are joinable, as in term rewriting systems. However, as
explained in the introduction, even a finite polygraphic rewriting system might
admit an infinite number of critical pairs. In the next section, we introduce a
theoretical setting which allows us to compute a finite number of generating
families of critical pairs.

\section{An embedding in compact 2-categories}
\label{sec:compact}
The notion of adjunction in the $2$-category~$\Cat$ of categories, functors and
natural transformations can be generalized to any $2$-category as
follows. Suppose that we are given a $2$-category~$\C$. A $1$-cell $f:A\to B$ is
\emph{left adjoint} to a $1$-cell~$g:B\to A$ (or $g$ is \emph{right adjoint}
to~$f$) when there exist two $2$-cells~$\eta:\id_A\To f\otimes g$
and~$\varepsilon:g\otimes f\To\id_B$, called respectively the \emph{unit} and
the \emph{counit} of the adjunction and depicted respectively on the left
of~\eqref{eq:adj}, such that $(f\otimes\varepsilon)\circ(\eta\otimes f)=\id_f$
and $(\varepsilon\otimes g)\circ(g\otimes\eta)=\id_g$. These equations are
called the \emph{zig-zag laws} because of their graphical representation, given
on the right of~\eqref{eq:adj}:
\begin{equation}
  \label{eq:adj}
  \strid{adj_unit}
  \qquad\quad
  \strid{adj_counit}
  \qquad\quad
  \strid{adj_zz_f_l}
  \qeq
  \strid{adj_zz_f_r}
  \qquad\qquad
  \strid{adj_zz_g_l}
  \qeq
  \strid{adj_zz_g_r}
\end{equation}
A $2$-category is \emph{compact} (sometimes also called \emph{autonomous} or
\emph{rigid}) when every $1$-cell admits both a left and a right adjoint. Given
a 2-category~$\C$, we write~$\fcomp{\C}$ for the free compact 2-category
on~$\C$. An explicit description of this 2-category can be
given~\cite{kelly-laplaza:coherence-compact-closed}:
\begin{itemize}
\item its 0-cells are the 0-cells of~$\C$,
\item its 1-cells are pairs~$f^n:A\to B$ consisting of an integer~$n\in\Z$,
  called \emph{winding number}, and a 1-cell $f:A\to B$ (\resp $f:B\to A$)
  of~$\C$ if $n$ is even (\resp odd),
\item a 2-cell is either~$\alpha^0:f^0\To g^0$, where $\alpha:f\To g$ is a
  2-cell of~$\C$, or $\eta_f^n:\id_B\To f^n\otimes f^{n+1}$ or
  $\varepsilon_f^n:f^{n+1}\otimes f^n\To\id_A$, where~$f^n:A\to B$ is a 1-cell,
  or a formal vertical or horizontal composite of those,
\item 1- and 2-cells are quotiented by a suitable congruence imposing the axioms
  of 2\nbd{}categories, compatibility of vertical and horizontal compositions
  in~$\fcomp\C$ with those of~$\C$ (for example
  $(\beta\circ\alpha)^0=\beta^0\circ\alpha^0$ and $(\id_f)^0=\id_{f^0}$) and the
  zig-zag laws~\eqref{eq:adj}.
\end{itemize}
Given a 1-cell~$f$ in this category, we often write~$f^m$ for the 1-cell which
is defined inductively by \hbox{$(f\otimes g)^m=f^m\otimes g^m$}
and~$(f^n)^m=f^{n+m}$ (notice that~$f^{-1}$ does not denote the inverse of~$f$
in this context). This algebraic construction is important in order to formally
define the 2\nbd{}category~$\fcomp\C$ but this construction might be better
grasped graphically, with the help of string diagrams: the compact structure
adds to~$\C$ the possibility to bend wires, without creating loops. For example,
consider a 2-cell \hbox{$\alpha:f\otimes g\To h\otimes i$} in a
2-category~$\C$. This 2-cell can be seen as a 2-cell \hbox{$\alpha^0:f^0\otimes
  g^0\To h^0\otimes i^0$} of~$\fcomp\C$, as pictured in the center
of~\eqref{eq:rotation}.
\begin{equation}
  \label{eq:rotation}
  \strid{rot_l}
  \qquad\qquad\qquad
  \strid{rot}
  \qquad\qquad\qquad
  \strid{rot_r}
\end{equation}
From this morphism, we can deduce a 2-cell $\rho_{f^0,g^0,h^0\otimes
  i^0}(\alpha):f^0\To h^0\otimes i^0\otimes g^1$, pictured on the right
of~\eqref{eq:rotation}, defined by~$\rho_{f^0,g^0,h^0\otimes
  i^0}(\alpha)=(\alpha\otimes\id_{g^1})\circ(\id_{f^0}\otimes\eta_g^0)$: the
wire corresponding to~$g^0$ can be bent on the right and the winding number is
increased by one (the output is of type~$g^1$) to ``remember'' that we have bent
the wire once on the right. Similarly, one can define from~$\alpha$ the
morphism~$\rho'_{f^0\otimes g^0,i^0,h^0}(\alpha):f^0\otimes g^0\otimes i^{-1}\To
h^0$, which corresponds to bending the wire of type~$i^0$ on the left, so its
winding number is decreased by $1$ (similar transformations can be defined for
bending the wires of type~$f^0$ and~$h^0$ in~$\alpha$). Interestingly, by the
definition of adjunctions, these two transformations provide mutual inverses:
\hbox{$\rho_{f,g,h}^{-1}=\rho'_{f,g,h}$}. We call
\emph{rotations} these bijections between the hom-categories of~$\fcomp\C$.

\begin{remark}
  \label{rem:border}
  The notions of source and target of a 2-cell in a compact 2-category is really
  artificial since, given a pair of parallel 1-cells $f,g:A\to B$, the rotations
  induce a bijection between the hom-categories~$\C(f,g)$
  and~$\C(\id_B,f^{-1}\otimes g)$.
\end{remark}

It can be shown that the winding numbers on the 1-cells provide enough
information about the bending of wires, so that
\begin{proposition}
  \label{prop:compact-embedding}
  Given a 2-category~$\C$, the embedding functor $E:\C\to\fcomp\C$ defined as
  the identity on 0-cells, as $f\mapsto f^0$ on 1-cells and as
  $\alpha\mapsto\alpha^0$ on 2-cells is full and faithful.
\end{proposition}
\noindent This means that given two 0-cells~$A$ and~$B$ of~$\C$, the
hom-categories~$\C(A,B)$ and~$\fcomp\C(A,B)$ are isomorphic in a coherent
way. The 2-category~$\fcomp\C$ thus provides a ``larger world'' in which we can
embed the 2-category~$\C$ without losing information.

The interest of this embedding is that there are ``extra morphisms''
in~$\fcomp\C$ that can be used to represent ``partial compositions''
in~$\C$. For example, consider two 2-cells~$\alpha:f\To f_1\otimes g\otimes f_2$
and $\beta:h_1\otimes g\otimes h_2\To h$ in~$\C$. These can be seen as the
morphisms of~$\fcomp\C$ depicted on the left of~\eqref{eq:partial-comp} by the
previous embedding.
\begin{equation}
  \label{eq:partial-comp}
  \strid{pcomp_a}
  \qquad\qquad
  \strid{pcomp_b}
  \qquad\qquad
  \strid{pcomp_rot}
  \qquad\qquad
  \strid{pcomp}
\end{equation}
From these two morphisms, the morphism \hbox{$\alpha\otimes_g\beta:f^0\To
  f_1^0\otimes h_1^{-1}\otimes h^0\otimes h_2^1\otimes f_2^0$}, depicted in the
center right of~\eqref{eq:partial-comp}, can be constructed. This morphism
represents the \emph{partial composition} of the 2-cells~$\alpha$ and~$\beta$ on
the 1-cell~$g$: up to rotations, this 2-cell is fundamentally a way to give a
precise meaning to the diagram depicted on the right of~\eqref{eq:partial-comp}.

The notion of 2-polygraph can easily be adapted to generate compact 2-categories
instead of 2-categories. Instead of generating a free category from the
underlying 1-polygraph, we generate a free category with winding numbers: with
the notations of Section~\ref{sec:polygraphs}, its objects are the elements
of~$E_0$ and its morphisms~$f_1^{n_1}\cdot f_2^{n_2}\cdots f_k^{n_k}:A\to B$ are
the paths $e(f_1^{n_1})\cdot e(f_2^{n_2})\cdots e(f_k^{n_k}):A\to B$ in the
graph described by the 1-polygraph, the edge~$e(f^n)$ being~$f$ is~$n\in\Z$ is
even or~$f$ taken backwards if~$f$ is odd. Similarly, instead of generating a
2-category from the polygraph, we generate a free compact 2-category on the
previously generated category with winding numbers with the 2-generators given
by the 2-polygraph. Such ``polygraphs'' are called \emph{compact polygraphs} and
we write~$\nCPol{2}$ for the category of compact 2-polygraphs. The embedding
given in Proposition~\ref{prop:compact-embedding} can be extended into an
embedding of~$\nPol{2}$ into~$\nCPol{2}$: every 2-polygraph can be seen as a
compact 2-polygraph. Given a compact 2-polygraph~$S$, the definition given in
Section~\ref{sec:cp} can be adapted in order to define the multicategory of
\emph{compact contexts}~$\contexts{S}$ of~$S$. Finally, the construction of nets
given in Section~\ref{sec:free-cat} can also be adapted in order to give a
concrete and implementable description of the multicategory~$\contexts{S}$ --
this essentially amounts to suitably adding winding numbers to 1-cells in the
polygraphs involved.

\begin{wrapfigure}{l}{16mm}
  \vspace{-4ex}
  \begin{center}
    $\strid{merge}$
  \end{center}
  \vspace{-4ex}
\end{wrapfigure}
Interestingly, the setting of compact contexts provides a generalization of
partial composition by allowing a ``partial composition of a morphism with
itself''. Namely, from a context
\hbox{$\alpha:(\ldots,(f_i,g_i),\ldots)\to(f,g^1\otimes h\otimes g^0)$} with
\hbox{$f:A\to A$} and \hbox{$h:B\to B$} one can build the context depicted on
the left~$\varepsilon_g^0\circ(g^1\otimes X\otimes
g^0)\circ\alpha:(\ldots,(f_i,g_i),\ldots,(h,\id_B))\to(f,\id_A)$, where
\hbox{$X:h\to\id_B$} is a fresh variable. This operation amounts to
\emph{merging} the outputs of type~$g^1$ and~$g^0$ of~$\alpha$.

\vspace{-2ex}
\section{The unification algorithm}
\label{sec:unification}
Now that the theoretical setting has been established, we can describe our
unification algorithm. Suppose that we are given a polygraphic rewriting system
$R\in\nPol{3}$ whose underlying signature is~$S=U_2(R)$. By the previous
remarks, $S$~can be seen as a compact 2-polygraph~$\fcomp{S}$. Now, suppose
that~$r_1$ and~$r_2$ are two rewriting rules (\ie 3-generators) in~$R$ whose
left member are respectively 2-cells~$\alpha:f\To g$ and~$\beta:h\To i$. The
2-cell~$\alpha:f\To g$ in the 2-category generated by~$S$ can be seen as a
2-cell~$\alpha^0:f^0\To g^0$ in the compact 2\nbd{}category~$\C$ generated
by~$\fcomp{S}$, and therefore as a nullary context~$\alpha:()\to(f^0,g^0)$ in
the multicategory of contexts $\contexts\C$. Similarly, $\beta$ can be seen as a
context~$\beta:()\to(h^0,i^0)$. In the multicategory $\contexts\C$, we can
compute a most general unifier of~$\alpha$ and~$\beta$ (see
Remark~\ref{rem:unif-contexts}) from which we will be able to generate critical
pairs of the rules~$r_1$ and~$r_2$. Because of space limitations, we don't
provide here a fully detailed and formal presentation of the algorithm: the
purpose of this paper was to introduce the formal framework necessary to define
the algorithm, whose in-depth description will be given in subsequent works.

We first introduce some terminology and notations on nets.
Given a 2-net~$\alpha$, an instance of a 2-generator~$y$ is the \emph{father}
(\resp \emph{son}) of an instance of a 1-generator~$x$ if~$x$ occurs in the
target (\resp source) of~$y$. For example, in~\eqref{eq:cnet}, $\gamma_0$ is a
son of~$1_0$ and~$1_1$ and a father of~$1_2$ and~$1_3$. It is easy to show that
a given instance of a 1-generator admits at most one father and one son. An
instance of 1-generator is \emph{dangling} when it has no father or no son. An
instance of a generator is in the \emph{border} of a net if it is in its source
or its target.

The algorithm proceeds as follows. We suppose that we have represented the
2-cells~$\alpha$ and~$\beta$ as polygraphic 2-nets. Our goal is to construct a
2-net~$\omega$ together with two injective morphisms of labeled
polygraphs~$i_1:\alpha\to\omega$ and~$i_2:\beta\to\omega$ satisfying the
properties required for unifiers as reformulated in
Remark~\ref{rem:unif-contexts}. The algorithm is quite similar to the rule-based
formulation of the unification algorithm for terms~\cite{baader-nipkow:trat}. It
begins by setting~$\omega=\alpha$ and~$i_1=\id_\alpha$, and then iterates a
procedure that will progressively propagate the unification and make~$\omega$
grow, by adding cells to it, until it is big enough so that there exists an
injection~$i_2:\beta\to\omega$. The procedure which is iterated is
non-deterministic and the critical pairs will be obtained as the collection of
the results of the non-failed branches of computation. During the iteration two
sets are maintained, $T$ and~$U$, which both contains pairs $(x,x')$ consisting
of an $n$-cell~$x$ of~$\beta$ and an $n$-cell~$x'$ of~$\omega$ for some
integer~$n\in\{0,1,2\}$. The set $U$ (for Unified) contains the injection~$i_2$
which is being constructed: if~$(x,x')\in U$ and the branch succeeds then the
resulting map~$i_2:\beta\to\omega$ will be such that~$i_2(x)=x'$. The set~$T$
(as in Todo) contains the pairs~$(x,x')$ such that~$x$ is a cell of~$\beta$
which is to be unified with the cell~$x'$ of $\omega$.

Initially, $\omega=\alpha$, $U=\emptyset$ and~$T=\{(x,x')\}$, where~$x$ and $x'$
are instances of 2-generators in~$\beta$ and in~$\omega$ respectively, both
chosen non-deterministically. Then the algorithm iterates over the following
rules, updating the values of~$\omega$, $U$ and~$T$ by executing the first rule
which applies (updating a value is denoted with the symbol~$\becomes$).
\begin{itemize}
\item \emph{Duplicate.} If~$T=\{(x,x')\}\uplus T'$ with $(x,x')\in U$ then
  $T\becomes T'$.
\item \emph{Clash.} If~$(x,x')\in T$ and $(x,x'')\in U$ and $x'\neq x''$ then
  fail.
\item \emph{Typecheck.} If~$(x,x')\in T$ with $\ell(x)\neq\ell(x')$ then fail.
\item \emph{Propagate-0.} If~$T=\{(x,x')\}\uplus T'$, where~$x$ and~$x'$ are
  0-cells then\\
  \begin{rpar}
  $T\becomes T'$ and $U\becomes\{(x,x')\}\cup U$.
  \end{rpar}
\item \emph{Propagate-1.} If~$T=\{(x,x')\}\uplus T'$, where~$x$ and~$x'$ are
  1-cells, then\\
  \begin{rpar}
  $T\becomes T'$ and\\
  if~$x$ has a father~$y$ then\\
  \begin{rpar}
    if~$x'$ has a father~$y'$ then\\
    \begin{rpar}
      $T\becomes\{(y,y')\}\cup T$ and $U\becomes\{(x,x')\}\cup U$
    \end{rpar}
    else either\\
    \begin{rpar}
      \null
      \begin{rpar}
        add a fresh generator~$y'$ of type~$\ell(y)$ in $\omega$,\\
        $T\becomes\{(y,y')\}\cup T$ and $U\becomes\{(x,x')\}\cup U$
      \end{rpar}
      or\\
      \begin{rpar}
        merge~$x'$ with some other 1-cell~$x''$ in the border of~$\omega$ in
        $\omega$,\\
        $T\becomes\{(x,x')\}\cup T$
      \end{rpar}
    \end{rpar}
  \end{rpar}
  if~$x$ has a son~$y$ then\\
  \begin{rpar}
    \emph{similar to the previous case}.
  \end{rpar}
  \end{rpar}
\item \emph{Propagate-2.} If~$T=\{(x,x')\}\uplus T'$, where~$x$ and~$x'$ are
  2-cells, then\\
  \begin{rpar}
    $T\becomes T'$, $U\becomes \{(x,x')\}\cup U$, we add in~$T$ that the 0- and
    1-cells in the source of~$x$ should be matched with the corresponding cells
    in the source of~$x'$, and the 0- and 1-cells of the target of~$x$ should be
    matched with those in the target of~$x'$.
  \end{rpar}
\end{itemize}
The ``either\ldots or'' construction above denotes a non-deterministic choice
and the ``merge'' refers to the merging operation introduced in
Section~\ref{sec:compact} (this operation might fail if the labels or the
winding numbers of~$x'$ and~$x''$ are not suitable).

The way this algorithm works is maybe best understood with an example. Consider
the signature~$S$ with one 0-cell $*$, one 1-cell $1:*\to *$ and three 2-cells
$\delta:1\to 4$, $\mu:4\to 1$ and $\sigma:1\to 1$ (where $4$ denotes $1\otimes
1\otimes 1\otimes 1$). We
write~$\varsigma=\sigma\otimes\sigma\otimes\sigma\otimes\sigma$. Now, consider a
rewriting system on this signature containing two rules~$r_1$ and~$r_2$ whose
left members are respectively~$\alpha=\varsigma\circ\delta$ and
$\beta=\mu\circ\varsigma$, that we represent respectively as the compact nets
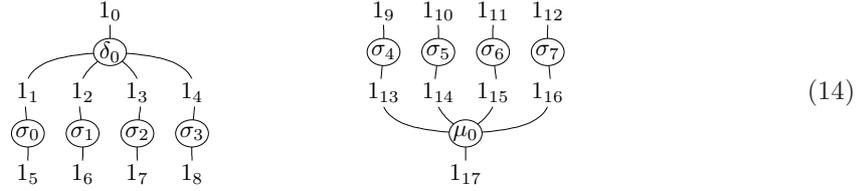
\begin{equation}
  \strid{unif_a}
  \qquad\qquad
  \strid{unif_b}
\end{equation}
(for simplicity, we omitted the instances of 0-cells).
We describe here a few possible non-determinis\-tic branches of the execution of
the algorithm. For example, if we begin with $T=\{(\sigma_4,\delta_0)\}$, the
algorithm will immediately fail by Typecheck because the label~$\sigma$
of~$\sigma_4$ differs from the label~$\delta$ of~$\delta_0$. Consider another
execution beginning with $T=\{(\sigma_4,\sigma_0)\}$, this time the label
matches so Propagate-2 will propagate the unification by setting
\hbox{$T=\{(1_9,1_1),(1_{13},1_5)\}$} and
$U=\{(\sigma_4,\sigma_0)\}$. Since~$1_9$ is dangling, Propagate-1 will move the
pair $(1_9,1_1)$ from~$T$ to~$U$. Then the pair $(1_{13},1_5)$ will be handled
by Propagate-1. Since~$1_5$ is dangling but~$1_{13}$ is not, a new
generator~$\mu_1$ will be added to~$\omega$ (now pictured on the left
of~\eqref{eq:unif-ex}) and after a few propagations $(1_{13},1_5)$ will be moved
from~$T$ to~$U$, $(\mu_0,\mu_1)$ will be added to~$U$ and~$T$ will
contain~$(1_{11},1_{19})$. By Propagate-1, this unification pair can lead to
multiple non-deterministic executions: a new generator~$\sigma_5$ can be added
(in the middle of~\eqref{eq:unif-ex}), or the 1-generator~$1_{19}$ can be merged
with another 1-generator ($1_7$ for example as pictured in the right
of~\eqref{eq:unif-ex}). Notice that in this last case, the morphism contains a
``hole'' of type~$1_6\To 1_{18}$, which is handled by a context variable.
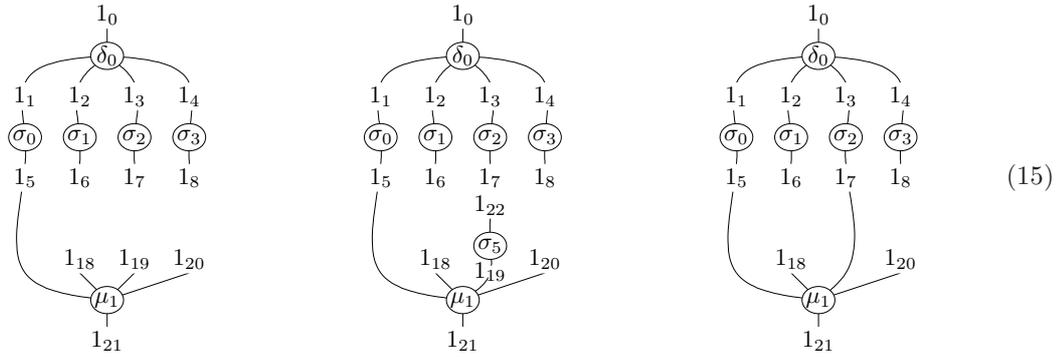
\begin{equation}
  \label{eq:unif-ex}
  \strid{unif_ex1}
  \qquad\qquad
  \strid{unif_ex2}
  \qquad\qquad
  \strid{unif_ex3}
\end{equation}
By executing fully the algorithm, the three morphisms of~\eqref{eq:unif-ex-res}
will be obtained as unifiers (as well as many others).
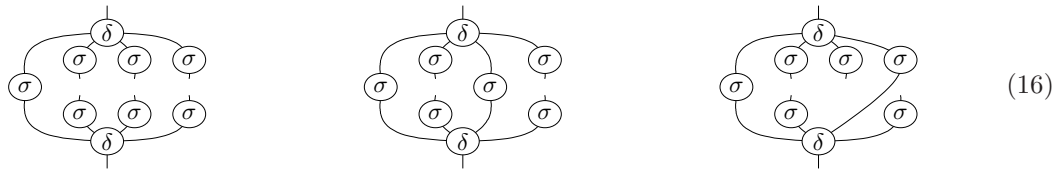
\begin{equation}
  \label{eq:unif-ex-res}
  \strid{unif_ex_res1}
  \qquad\qquad
  \strid{unif_ex_res2}
  \qquad\qquad
  \strid{unif_ex_res3}
\end{equation}

It can be shown that the algorithm terminates and generates all the critical
pairs in compact contexts, and these are in finite number. It is important to
notice that the algorithm generates the critical pairs of a rewriting system~$R$
in the ``bigger world'' of compact contexts, from which we can generate the
critical pairs in the 2-category generated by~$R$ (which are not necessarily in
finite number as explained in the introduction). If joinability of the critical
pairs in compact contexts implies that the rewriting system is confluent, the
converse is unfortunately not true: a similar situation is well known in the
study of $\lambda$-calculus with explicit substitution, where a rewriting system
might be confluent without being confluent on terms with metavariables.

We have realized a toy implementation of the algorithm in less than 2000 lines
of OCaml, with which we have been able to successfully recover the critical
pairs of rewriting systems in~\cite{lafont:boolean-circuits}. Even though we did
not particularly focus on efficiency, the execution times are good, typically
less than a second, because the morphisms involved in polygraphic rewriting
systems are usually small (but they can generate a large number of critical
pairs)

\paragraph{\textbf{Future works.}}
This paper lays the theoretical foundations for unification in polygraphic
2\nbd{}dimensional rewriting systems and leaves many research tracks open for
future works. We plan to study the precise links between our algorithm and the
usual unification for terms (every term rewriting system can be seen as a
polygraphic rewriting system~\cite{burroni:higher-word}) as well as algorithms
for (planar) graph rewriting. Concerning concrete applications, since these
rewriting systems essentially transform circuits made of operators (the
2-generators) linked by a bunch of wires (the 1-generators), it would be
interesting to see if these methods can be used to optimize electronic
circuits. Finally, we plan investigating the generalization of these methods in
dimension higher than 2, which seems to be very challenging.

\paragraph{\textbf{Acknowledgements.}}
The author is much indebted to John Baez, Albert Burroni, Jonas Frey, Emmanuel
Haucourt, Martin Hyland, Yves Lafont, Paul-André Melliès and François Métayer.

\bibliographystyle{plain}
\bibliography{these}
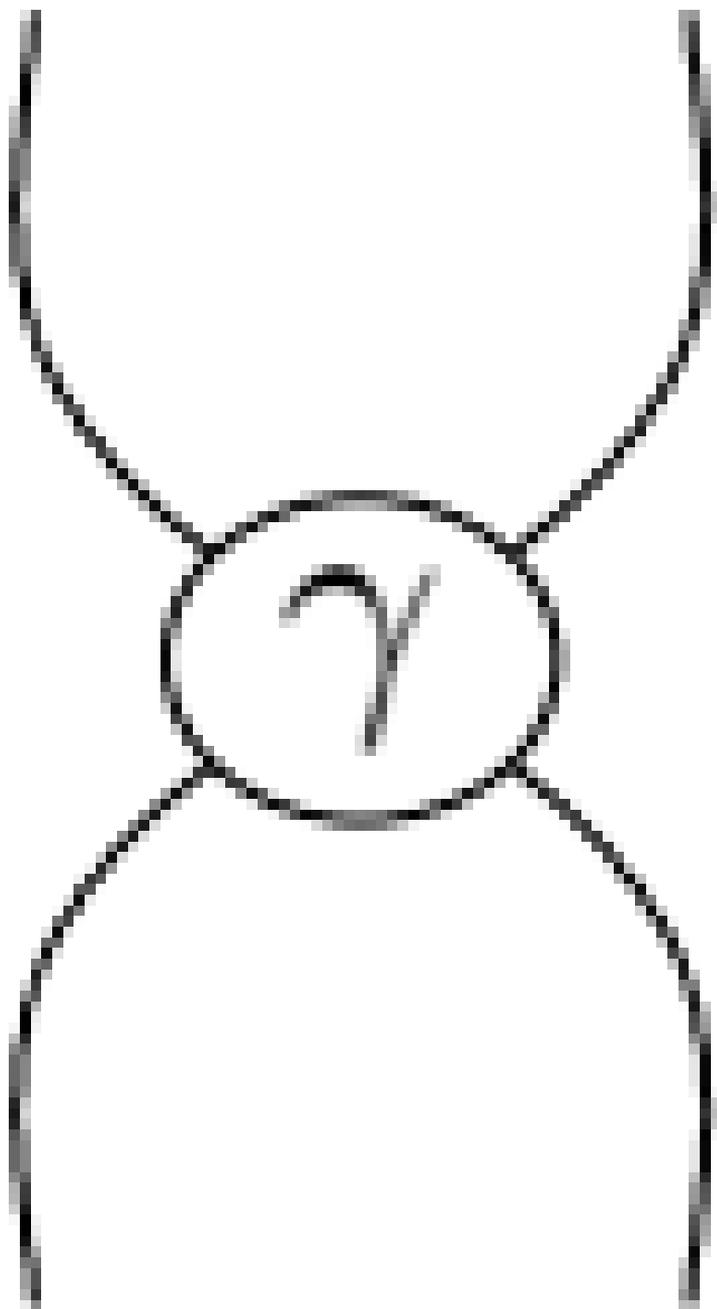
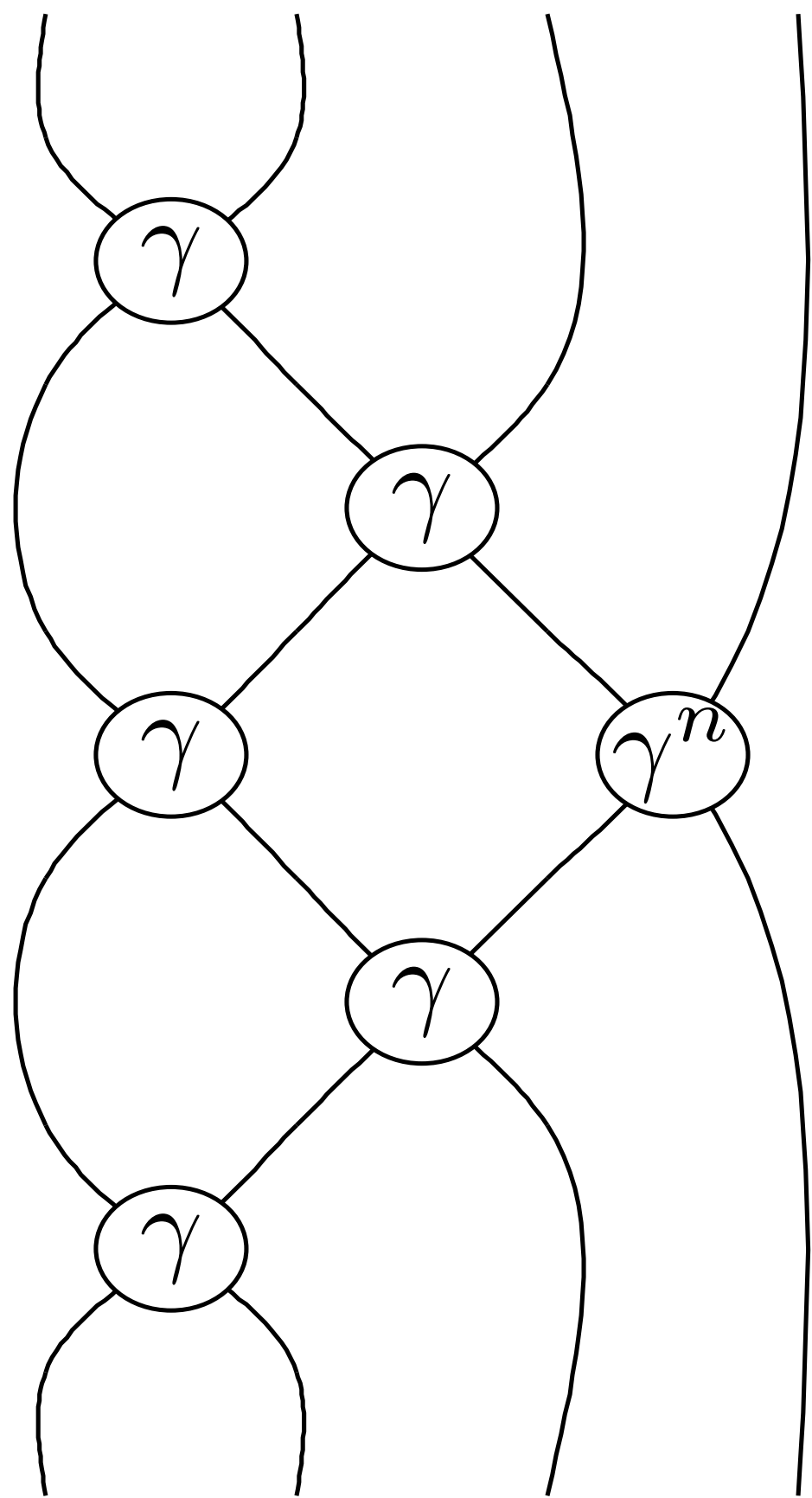
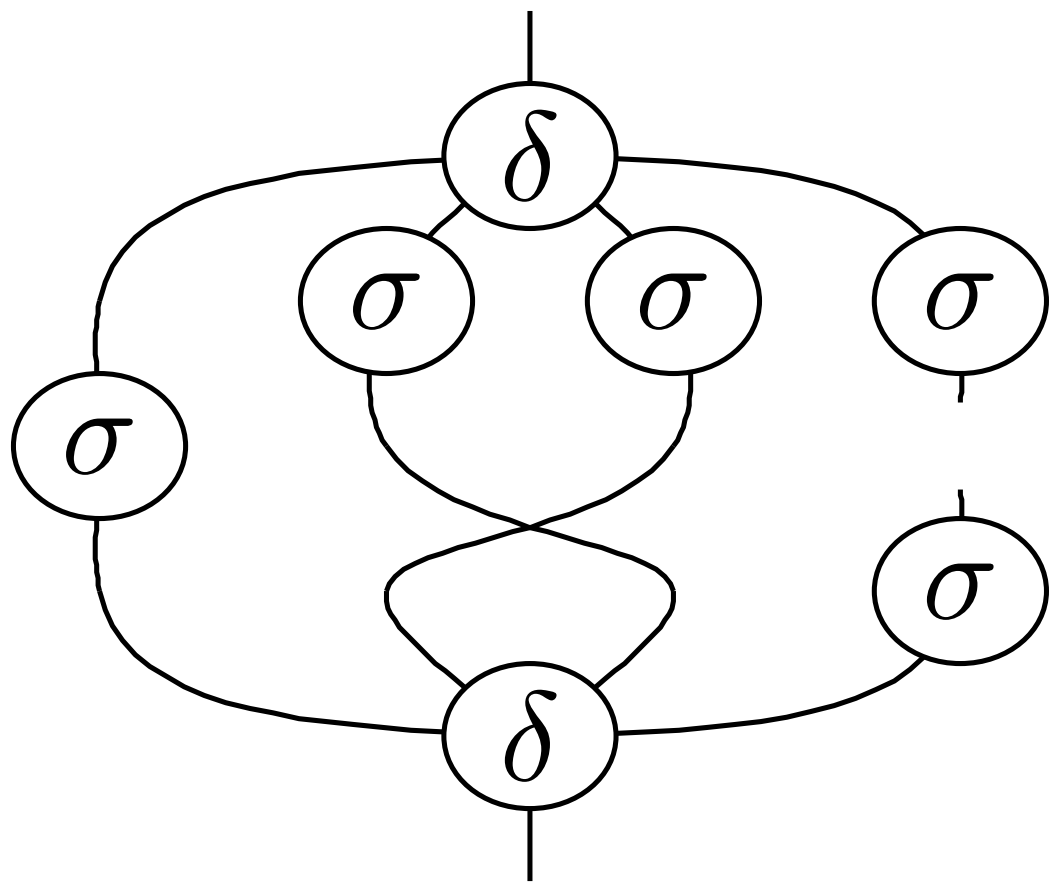
\end{document}